\documentclass[11pt,a4paper]{article} 
\pdfoutput=1
\usepackage{jcappub}
\usepackage{latexsym,amsmath,amssymb}
\usepackage{graphicx}
\usepackage{slashed}  
\usepackage{bm}
\usepackage{epsfig}
\usepackage{dcolumn}

\newcommand{\beq}{\begin{equation}}
\newcommand{\eeq}{\end{equation}}
\newcommand{\bea}{\begin{eqnarray}}
\newcommand{\eea}{\end{eqnarray}}
\newcommand{\p}{\partial}
\newcommand{\ep}{\epsilon}
\newcommand{\lle}{\left<}
\newcommand{\rgr}{\right>}
\newcommand{\lb}{\left|}
\newcommand{\rb}{\right|}

\newcommand{\Hint}{H_{\mathrm{int}}}

\newcommand{\Torder}{\mathrm{T}}
\newcommand{\vect}[1]{\bm{\mathrm{{#1}}}}
\newcommand{\hf}{\frac{1}{2}}

\title{On the non-Gaussian correlation of the primordial curvature perturbation with vector fields}

\author[a,b]{Rajeev Kumar Jain}
\author[b]{and Martin S. Sloth}

\affiliation[a]{D\'epartement de Physique Th\'eorique and Center for Astroparticle Physics, \\
Universit\'e de Gen\`eve, 24, Quai E. Ansermet, CH--1211 Gen\`eve 4, Switzerland}
\affiliation[b]{CP$^3$-Origins, Centre for Cosmology and Particle Physics Phenomenology, \\
University of Southern Denmark, Campusvej 55, 5230 Odense M, Denmark}

\emailAdd{rajeev.jain@unige.ch}
\emailAdd{sloth@cp3.dias.sdu.dk}

\date{\today}

\abstract{
We compute the three-point cross-correlation function of the primordial curvature perturbation generated during inflation with two powers of a vector field in a model where conformal invariance is broken by a direct coupling of the vector field with the inflaton. If the vector field is identified with the electromagnetic field, this correlation would be a non-Gaussian signature of primordial magnetic fields generated during inflation. We find that the signal is maximized for the flattened configuration where the wave number of the curvature perturbation is twice that of the vector field and in this limit, the magnetic non-linear parameter becomes as large as $|b_{NL}| \sim \mathcal O(10^{3})$.  In the squeezed limit where the wave number of the curvature perturbation vanishes, our results agree with the {\it magnetic consistency relation} derived in arXiv:1207.4187.}

\keywords{Inflation, primordial magnetic fields, non-Gaussianity}
\arxivnumber{}

\begin{document}
\maketitle


\section{Introduction}

The observations of the acoustic peaks in the Cosmic Microwave Background (CMB) and the underlying nearly flat spectrum of primordial curvature perturbations are so far the main observational verification of the inflationary paradigm and has ruled out some of the leading competing scenarios such as cosmic string structure formation \cite{Lyth:1998xn}. Despite the success of the inflationary paradigm any firm knowledge about the underlying microscopic model is still lacking. In fact, even the energy scale of inflation is still unknown within ten orders of magnitude, but a possible detection of primordial gravitational waves in the future would be able to pin it down. Similarly a detection of primordial non-Gaussianity or isocurvature modes would give us more detailed information about the underlying model of inflation. These are some of the new observables, that we might be fortunate to probe with Planck and other present and future experiments, but it is strongly model dependent if these observables will turn out to be within experimental reach \cite{planck}. It is therefore important to keep looking for new observables, which could be significant in other non-minimal versions of inflation.

On the other hand astrophysical observations indicate the existence of coherent magnetic fields on different cosmological scales. The strength of magnetic fields in galaxies and galaxy clusters has been determined to be of the order of a micro Gauss \cite{Kronberg:1993vk}, while a lower bound of $0.1$ femto Gauss on magnetic fields in the intergalactic medium has also been claimed \cite{Neronov:1900zz,Taylor:2011bn,Vovk:2011aa}. While astrophysical processes may be disfavored for generating coherent magnetic fields on such large scales, inflation by its very nature creates coherent fluctuations on very large scales. It is therefore natural to speculate that large scale primordial magnetic fields could be generated during inflation by a coupling between the inflaton field and the vector field, which breaks the conformal invariance of the vector field during inflation \cite{Ratra:1991bn}. As we discuss below, such models of primordial magnetogenesis suffer from the strong coupling problem \cite{Demozzi:2009fu}. However, if gauge fields (which may or may not be identified with the electromagnetic field) are produced this way, they will be correlated with the primordial curvature perturbation generated during inflation. 

Motivated by these observations, the focus of the present paper is therefore to compute the three-point cross-correlation function between the comoving curvature perturbation and two powers of the gauge field (to be identified with the electromagnetic field in models of magnetogenesis). We have recently derived a {\it magnetic consistency relation} \cite{Jain:2012ga}, which can be used to check the results in the so-called squeezed limit, and we find very good agreement between the consistency relation and the results presented here.

Related calculations, which have partly served, as motivation for the present work, have previously been carried out in \cite{Caldwell:2011ra} in the more restricted setting of a test scalar field e.g. curvaton \cite{Enqvist:2001zp,Lyth:2001nq,Moroi:2001ct} coupled to electromagnetism. Subsequently, the cross-correlation of the comoving curvature perturbation with the magnetic fields in a specific model was also considered \cite{Motta:2012rn}. Our calculations can be regarded as a generalization and check of these results, although there are some disagreements, as we will discuss later.

The time-dependent coupling of the $U(1)$ gauge field to the background (or the scalar field) can be parametrized by a coupling of the form $\lambda(\phi)F_{\mu\nu}F^{\mu\nu}$, where $F_{\mu\nu} \equiv \p_\mu A_\nu-\p_\nu A_\mu$ is the field strength associated to the $U(1)$ gauge field, and the time dependence of the coupling is parametrized by the dependence on some slowly rolling background field $\phi$, which we will think of as being the inflaton for simplicity. If the vector field is identified with the electromagnetic field this is a well studied model of primordial magnetogenesis  \cite{Ratra:1991bn,Turner:1987bw,Widrow:2002ud}. It has been proposed that such type of models could yield the required magnetic seed fields for the galactic dynamo \cite{Bamba:2003av,Bamba:2006ga,Martin:2007ue,Kanno:2009ei,Subramanian:2009fu,Durrer:2010mq,Byrnes:2011aa,Jain:2012jy,Kahniashvili:2012vt,Urban:2012ib}, and perhaps lead to signatures in the CMB \cite{Kosowsky:1996yc,Lewis:2004ef,Lewis:2004kg,Kosowsky:2004zh,Giovannini:2008aa,Kahniashvili:2008hx,Kristiansen:2008tx,Seshadri:2009sy,Caprini:2009vk,Cai:2010uw,Shiraishi:2010kd,Brown:2010jd,Giovannini:2009zq,Yamazaki:2010nf,Kahniashvili:2010wm,Bonvin:2010nr,Durrer:1999bk,Paoletti:2008ck,Finelli:2008xh,Giovannini:2008yz,Kahniashvili:2006hy,Mack:2001gc,Paoletti:2010rx,Paoletti:2012bb,Shaw:2010ea,Subramanian:1997gi,Yamazaki:2012pg}. Even if the vector field is not identified with the electromagnetic field, it has previously been argued that a significant signal of non-Gaussianity \cite{Barnaby:2012tk} and statistical anisotropy (for the power spectrum and the bispectrum) might be a natural outcome in such models where a vector field is present during inflation \cite{Yokoyama:2008xw,Bartolo:2012sd}.

We would like to emphasize that there are some theoretical obstacles for a model of this type to succeed as a working scenario of primordial magnetogenesis, when the gauge field is associated with the electromagnetic field. In order to produce significant magnetic fields without running into problems with large back reaction on large or small scales, the spectrum of the magnetic fields has to be nearly scale invariant. On the other hand, a scale invariant spectrum of magnetic fields can only be obtained for $n=2$ and $n=-3$, when the coupling $\lambda(\phi) \sim a^n(t)$ with $a(t)$ being the scale factor, varies by many orders of magnitude during inflation. This implies that for $n=2$, if $\lambda(\phi)$ at the end of inflation is unity, i.e., $\lambda_I=1$, as required to obtain standard electromagnetism, the effective gauge coupling squared, scaling inversely with $\lambda$, must have been very large at the beginning of inflation and one would have started out in a regime of strong coupling \cite{Demozzi:2009fu}. On the other hand, when $n =-3$, which is in the weakly coupled regime, it has been shown that the energy density in the electromagnetic field scales as $\varepsilon_{EM}\sim H_I^4 a^{-2(n+2)}$. Therefore, for backreaction to remain small, i.e., $\varepsilon_{EM} \ll H_I^2$, for a typical value of $H_I\sim 10^{-5}$, we can maximally obtain only about $12$ e-folds of inflation before backreaction becomes important, which is much less than the required number of e-folds. 

As discussed in \cite{Barnaby:2012tk} it is very difficult to see how the strong coupling issue can be avoided retaining gauge invariance, unless one gives up on the relation to primordial magnetogenesis and on the gauge field being the actual electromagnetic field, and instead thinks of it as some novel hidden sector isocurvature gauge field. Another possibility, which has been proposed, is that the gauge invariance is broken in the $UV$ in the effective four dimensional Lagrangian, but manifest in five dimensions \cite{Motta:2012rn}. If gauge invariance is broken, the effective gauge coupling squared may not scale inversely with $\lambda(\phi)$. It has also been proposed that the gauge field could instead be identified with a vector curvaton, in which case the constraint on the gauge coupling at the end of inflation doesn't apply. In this case we would have a new type of non-Gaussianity in a scenario of mixed inflation and vector curvaton perturbations \cite{Jain:2012ga}. Here we will be agnostic about the detailed role of the gauge field, and leave further exploration of possible applications of the model for future work. However, for the remaining of this paper, we will for definiteness refer to the gauge field as if it is the electromagnetic field, which will also make the comparison with the previous work in \cite{Caldwell:2011ra,Motta:2012rn} more direct. 

It is not immediately obvious what is the most convenient way to parametrize the results. If we define the cross-correlation bispectrum of the curvature perturbation with the magnetic fields as\footnote{In certain physical applications, as when comparing to the induced $f_{NL}$ in the CMB, it may be convenient to straightforwardly symmetrize this expression over $k_1$, $k_2$, and $k_3$.}
\beq \label{zBB0}
\left<\zeta(\vect{k}_1){\bf B}(\vect{k}_2)\cdot {\bf B}(\vect{k}_3)\right> \equiv(2\pi)^3\delta^{(3)}(\vect{k}_1+\vect{k}_2+\vect{k}_3)B_{\zeta B B}(\vect{k}_1,\vect{k}_2, \vect{k}_3) ~,
\eeq
then as we have previously proposed, it is convenient to define the magnetic non-linearity parameter\footnote{Note, a different dimensionless quantity  $B_{\zeta BB}(\vect{k}_1,\vect{k}_2, \vect{k}_3)/(\sqrt{P_\zeta(k_1) P_B(k_2)P_B(k_3)})$ was previously introduced in \cite{Caldwell:2011ra,Motta:2012rn}, which is different from our $b_{NL}$. Expressing the non-linearities through $b_{NL}$, rather than through previously introduced quantities, makes the understanding of the induced non-Gaussianity in the CMB and large scale structure more straightforward.} $b_{NL}$, in terms of the cross-correlation function of the curvature perturbation with the magnetic fields
\beq \label{zBB1}
B_{\zeta B B}(\vect{k}_1,\vect{k}_2, \vect{k}_3) \equiv b_{NL}P_\zeta(k_1) P_B(k_2)~,
\eeq
where $P_\zeta$ and $P_B$ are the power spectra of the comoving curvature perturbation and the magnetic fields, 
defined respectively as  
\bea
\left<\zeta(\vect{k})\zeta(\vect{k}')\right> \equiv(2\pi)^3\delta^{(3)}(\vect{k}+\vect{k}') P_\zeta(k),\\
\left<{\bf B}(\vect{k})\cdot {\bf B}(\vect{k}')\right> \equiv(2\pi)^3\delta^{(3)}(\vect{k}+\vect{k}') P_B(k).
\eea

In the case where $b_{NL}$ is momentum independent, it takes a ``local" form which can be derived from the relation
\beq
{\bf B} = {\bf B}^{(G)} +\frac{1}{2}b_{NL}^{local} \zeta^{(G)}  {\bf B}^{(G)}
\eeq
with ${\bf B}^{(G)}$ and $\zeta^{(G)}$ being the Gaussian fields.
There is an interesting limit where the magnetic non-linearity parameter takes the local form, which makes the comparison with the above estimate particularly simple. We will show that in the squeezed limit, where the  momentum of the curvature perturbation vanishes, i.e., $k_1\ll k_2,k_3=k$, we, in fact, recover 
\beq
\left<\zeta(\vect{k}_1){\bf B}(\vect{k}_2)\cdot {\bf B}(\vect{k}_3)\right> =b_{NL}^{local}(2\pi)^3\delta^{(3)}(\vect{k}_1+\vect{k}_2+\vect{k}_3)P_\zeta(k_1) P_B(k)~, 
\eeq
with $ b_{NL}^{local} = n_B-4$ where $n_B$ is the spectral index of the magnetic field power spectrum, in agreement with the magnetic consistency relation, which was derived in \cite{Jain:2012ga} using simpler semi-classical methods\footnote{This approach is a non-trivial generalization of related semiclassical methods used in \cite{Maldacena:2002vr,Creminelli:2004yq,Seery:2008ax,Giddings:2010nc,Giddings:2011zd,Creminelli:2012ed,Kehagias:2012pd}. In the appendix of  \cite{Giddings:2010nc}, some of these approaches are reviewed.}. In the case of a scale invariant spectrum of magnetic fields, $n_B= 0$, we have $b^{local}_{NL} = -4$. 

Another interesting limit which maximizes the three-point cross-correlation function is the flattened shape where $k_1/2=k_2=k_3$. In this limit it turns out that the signal is enhanced by a logarithmic factor in agreement with \cite{Caldwell:2011ra}. On the largest scales the logarithm will give an enhancement by a factor 60. Thus, for a flat magnetic field power spectrum, the non-linearity parameter in the flattened limit becomes $|b_{NL}|  \sim  \mathcal O(10^3) $ depending on the scale.

The paper is organized as follows. In the next section we will review the model and the mechanism for the production of large scale gauge fields during inflation, which has been proposed as a mechanism of primordial magnetogenesis. In section \ref{inthamil}, we derive the interaction Hamiltonian for the primordial curvature perturbation and the gauge field, while in section \ref{crosscorr} we calculate the three-point cross-correlation function of the curvature perturbation with two powers of the magnetic field. We show that $b_{NL}$ is maximal in the flattened shape and of order $\mathcal{O}(10^3)$. We then derive the squeezed limit of our result, and compare it with the magnetic consistency relation. Finally, in section \ref{diss}, we conclude with a discussion of our results. In the appendix we have listed some useful integrals.

Throughout this paper, we work in natural units with $\hbar = c =1$, and the Planck mass $M_p^2 \equiv 1/8\pi G$ equal to unity. Our metric convention is $(-,+,+,+)$.


\section{Primordial magnetogenesis} \label{pmf}

In order to break the conformal invariance of the Maxwell equations, a necessary condition for amplification of magnetic fields during inflation, we consider the coupling of the electromagnetic field with a scalar field \cite{Ratra:1991bn}
\beq
S=\frac{1}{2}\int d^4 x\sqrt{-g}\left(R-\p_{\mu}\phi\p^{\mu}\phi -2V(\phi)-\frac{1}{2}\lambda(\phi)F_{\mu\nu}F^{\mu\nu}\right)~.
\eeq	
In general, the scalar field doesn't need to be the inflaton, but could be any light field (or a set of light fields) in slow-roll during inflation. However, for simplicity we will assume by default that $\phi$ is the inflaton, and comment in the appropriate places how the result would change if $\phi$ is some other light field like a curvaton or an entropy perturbation mode. If $\phi$ is the inflaton field, it has to satisfy the usual slow-roll conditions
\beq
\ep \equiv \frac{1}{2}\left(\frac{V'}{V}\right)^2 \ll 1~,\qquad \eta \equiv \frac{V''}{V} \ll 1~.
\eeq 

We find it convenient to work in the Coulomb gauge with $A_0=0$ and $\p_i A^i=0$, where the quadratic action for the electromagnetic vector field, $A_i$, becomes
\beq\label{A2}
S_{\rm em} = -\frac{1}{4}\int d^4 x \sqrt{-g}\,\lambda(\phi) F_{\mu\nu} F^{\mu\nu}=\frac{1}{2} \int d^3 x\, d\tau \lambda(\phi) \left({A_i'}^2 -\frac{1}{2}(\p_iA_j-\p_jA_i)^2\right)~.
\eeq
The conformal time, $\tau$, and the scale factor, $a(\tau)$, are defined by the choice of the background metric $ds^2=a^2(\tau)(-d\tau^2+d\vect{x}^2)$.

We are interested in calculating correlation functions in Fourier space as in (\ref{zBB1}), where the Fourier transformed gauge field is 
\beq
A_i(\tau,\vect{k}) = \int d^3 x A_i(\tau,\vect{x}) e^{i \vect x \cdot \vect k}
\eeq
In order to quantize the gauge field, we define the usual mode expansion 
\beq
A_i(\tau,\vect{k}) 
=\sum_{\sigma = \pm} \left[\ep_i^{\sigma}(\hat k)A_k(\tau)\hat b^{\sigma}_{\vect{k}}e^{i\vect{k}\cdot \vect{x}} + h.c.\right]
\eeq
and impose the standard commutation relations
\beq
[\hat b^{\sigma}_{\vect{k}},\hat {b^{\sigma'}_{\vect{k}'}}^{\dagger}] =(2\pi)^3\delta^{(3)}(\vect{k}-\vect{k}')\delta_{\sigma\sigma'} ~.
\eeq
The polarization vectors will have to satisfy 
\beq
\vect{k}\cdot \ep^{\sigma}(\hat k) =0~,\quad  \ep^{\sigma}(\hat k)\cdot{\ep^{\sigma'}}^*(\hat k) =\delta_{\sigma \sigma'}~,\quad \sum_{\sigma=\pm} \ep_i^{\sigma}(\hat k){\ep_j^{\sigma}}^*(\hat k) =\delta_{ij} - k_i k_j/k^2~.
\eeq
For the two point correlation function, we then have
\beq
\left< A_i(\tau,\vect{k}) A_j(\tau,\vect{k}') \right>= (2\pi)^3 \delta^{(3)}(\vect{k}+\vect{k}')\left(\delta_{ij}-\frac{ k_i  k_j}{k^2}\right) \left| A_k(\tau)\right|^2~.
\eeq
We can now compute the magnetic field in terms of the gauge field, by using the relation 
\beq
B_i(\tau,\vect{x}) = \frac{1}{a}\ep_{ijk}\p_jA_k(\tau,\vect{x})~.
\eeq
Upon using $\ep_{ijl}\ep^{imn} = \delta_j^m\delta_l^n-\delta_j^n\delta_l^m$, we find that the two point correlation function of the magnetic fields is given by 
\bea
\left< B_i(\tau,\vect{k}){B^i}(\tau,\vect{k}')\right> &=& \frac{k^2}{a^4}\left(\delta_{ij}-\frac{ k_i k_j}{k^2}\right) \left< A_i(\tau,\vect{k}) A_j(\tau,\vect{k}') \right>\nonumber\\
&=&2\frac{k^2}{a^4}(2\pi)^3\delta^{(3)}(\vect{k}+\vect{k}')|A_k(\tau)|^2
\eea
If we define the magnetic power spectrum as
\beq
\left< B_i(\tau,\vect{k}){B^i}(\tau,\vect{k}')\right> =(2\pi)^3\delta^{(3)}(\vect{k}+\vect{k}') P_B(\tau,k)~,
\eeq
then the magnetic power spectrum is given by 
\beq
 P_B(\tau,k)=2\frac{k^2}{a^4}|A_k(\tau)|^2~.
\eeq

It is convenient to parametrize the time dependence of the coupling as $\lambda(\phi(\tau))= \lambda_I(\tau/\tau_I)^{-2n}$. For an explicit toy model of this type, one might consider inflation with a linear potential, and the coupling $\lambda$ as an exponential function of the inflation field \cite{Caldwell:2011ra}. Defining the pump field $S^2(\tau) =\lambda(\phi(\tau))$ and the canonically normalized vector field $v_i= S(\tau) A_i$, the quadratic action for the vector field takes the canonical form for the kinetic term
\beq
S_{\rm em} =   \frac{1}{2}\int d\tau d^3 x  \left[v_i'^2-(\p_j v_i)^2+\frac{S''}{S}v_i^2\right]~,
\eeq
and the equation of motion for the mode function, $v_k = S(\tau) A_k$, is
\beq
v_k'' +\left(k^2-\frac{S''}{S}\right)v_k=0~.
\eeq
The solution, normalized to the Bunch-Davies vacuum, is
\beq\label{vk}
v_k(\tau) = \frac{\sqrt{\pi}}{2}e^{i\pi(1+n)/2}\sqrt{-\tau}H^{(1)}_{\frac{1}{2}+n}(-k\tau)~,
\eeq
which leads to \cite{Caldwell:2011ra}
\beq
P_B(\tau,k) =\frac{1}{\lambda_I} \frac{\pi}{2}\frac{H^4}{k^3}\left(\frac{\tau}{\tau_I}\right)^{2n}(-k\tau)^5H^{(1)}_{\frac{1}{2}+n}(-k\tau)H^{(2)}_{\frac{1}{2}+n}(-k\tau)~.
\eeq
Here $H_{n}^{(1,2)}(x)$ are the Hankel functions of the first and second kind, $H$ is the Hubble parameter during inflation, $\tau_I$ is the conformal time at the end of inflation, and $\lambda_I$ is the coupling $\lambda(\tau)$ evaluated at the end of inflation at $\tau=\tau_I$.

Assuming a resolution to the strong coupling problem as in \cite{Motta:2012rn}, it can be estimated that for $n=2$, where backreaction remains small and with $H\simeq 10^{14}$\,GeV, a magnetic field strength of $\sim 1$ nano Gauss can be achieved on Mpc scales \cite{Caldwell:2011ra, Demozzi:2009fu}. 


\section{The interaction Hamiltonian} \label{inthamil}

To find the interaction Hamiltonian of the metric fluctuations with the electromagnetic fields, it is convenient to write the metric in the ADM form as
\beq
ds^2 = -N^2 dt^2 + h_{ij}(dx^i+N^i dt)(dx^j+N^jdt)~,
\eeq
where the dynamic degrees of freedom are contained in $h_{ij}$, and the lapse, $N$, and the shift, $N^i$, are determined by the constraint equations, which can be derived from the Lagrangian obtained by inserting the ADM decomposition of the metric into the background action
\bea\label{Lsigma}
\mathcal{L} &=& \frac{a^3}{2}\left[NR^{(3)} -2N V+ N^{-1}(E^j_i E^i_j-(E^i_i)^2)\right.\nonumber\\
 & &\left. +\, N^{-1}(\dot\phi-N^i\p_i\phi)^2
-N a^{-2}[e^{-\gamma}]^{ij}\p_i\phi\p_j\phi\right]~,
\eea
where $E_{ij} =\hf( \dot h_{ij} -\nabla_iN_j-\nabla_jN_i)$ is the rescaled extrinsic curvature, $V$ is the inflaton potential, and $R^{(3)}$ is the curvature scalar of the three-metric $h_{ij}$.  

The dynamical degrees of freedom can be parametrized by a scalar curvature perturbation, $\zeta$, and traceless and transverse tensor perturbation, $\gamma_{ij}$, by writing
\beq
h_{ij} = a^2 e^{2\zeta} \left[e^{\gamma}\right]_{ij}
\eeq
In this way, one obtains for the lapse and the shift to linear order
\bea\label{N}
N&=& 1+\frac{1}{H}\dot\zeta\nonumber\\
N_i&=&\p_i\left(-\frac{1}{H}\zeta+a^2 \ep\, \p^{-2}\dot\zeta\right)~,
\eea
Similarly quadratic action for the curvature perturbation becomes 
\beq
S= \int d^3 x\, d\tau\, a^2(\tau)\, \ep \left[\zeta'^2-(\p\zeta)^2\right]~.
\eeq
Expanding in terms of mode functions
\beq
\zeta(\tau,\vect{x}) =\int \frac{d^3 k}{(2\pi)^3}\left[a_{\vect{k}}\zeta_{k}(\tau)e^{i\vect{x}\cdot\vect{k}} +h.c.\right]
\eeq
the mode equation becomes
\beq
\zeta_k'' +2aH(1+\delta+\ep)\zeta_k'+k^2\zeta_k=0
\eeq
with $\ep=-\dot H/H^2$ and $\delta=\ddot H/2H\dot H$. The solution normalized to the Bunch-Davis vacuum to leading order in slow-roll is given by 
\beq\label{zeta}
\zeta_k(\tau) = \frac{1}{\sqrt{2\ep}}\frac{H}{\sqrt{2k^3}}(1+ik\tau)\,e^{-ik\tau}~.
\eeq

Under linear variation of the Lagrangian in the metric, the interaction Hamiltonian is given by
\beq\label{HzAA}
H_{\zeta AA} =-\frac{1}{2} \int d^3 x\, a^3\, T^{\mu\nu}\delta g_{\mu\nu}~.
\eeq
Upon using (\ref{N}) in (\ref{HzAA}), the interactions of the metric fluctuations with gauge fields at third order are given by \cite{Chaicherdsakul:2006ui}
\beq\label{Hzaa}
H_{\zeta AA} = \int d^3 x a^3\left(\frac{1}{H}\dot\zeta T^{00} -\p_i\left(-\frac{1}{H}\zeta +a^2 \ep \p^{-2}\dot\zeta\right)T^{0i}-a^2\zeta T^{ii}\right)~.
\eeq
Inserting directly the energy-momentum tensor of the gauge field
\beq
T_{\mu\nu} = \lambda F_{\mu\rho}{F_{\nu}}^{\rho}-\frac{1}{4}g_{\mu\nu}\lambda F_{\rho\kappa}F^{\rho\kappa}~,
\eeq
one finds an expression that agrees with equation (46) of \cite{Motta:2012rn} and is of order $\lambda\zeta A A$. However, the true interaction is actually of lower order since the interaction used in \cite{Motta:2012rn} is proportional to a total time derivative to leading order. Using that 
\beq
\nabla_{\mu} T^{\mu\nu} =-\frac{1}{4}(\nabla^{\nu}\lambda)F_{\rho\kappa}F^{\rho\kappa} ~,
\eeq
which in the comoving Coulomb gauge yields
\beq 
\nabla_{\mu} T^{\mu 0} = -\frac{1}{2a^2}\dot\lambda\left(\dot A_i\dot A_i -\frac{1}{2a^2}(\p_i A_j-\p_j A_i)^2\right)~,
\eeq
where the non-vanishing right-hand-side is due to the sourcing between the scalar and vector modes. The total energy-momentum tensor of the gauge field and the inflaton is of course conserved. 

Writing explicitly the covariant derivative in terms of the background metric, one finds
\beq\label{bianchi}
\frac{1}{a^3}\p_t \left(a^3 T^{00}\right) +a\dot a T^{ii} +\p_i T^{i0} +\frac{1}{2a^2}\dot\lambda \left(\dot A_i\dot A_i -\frac{1}{2 a^2}(\p_i A_j-\p_j A_i)^2\right)=0~,
\eeq
so clearly, by inserting (\ref{bianchi}) into (\ref{Hzaa}), we obtain after some partial integrations to leading order in slow-roll
\beq\label{Hint}
H_{\zeta AA} = \int d^3 x \left(a\dot\lambda\frac{1}{H}\zeta\left(\frac{1}{2}\dot A_i\dot A_i -\frac{1}{4 a^2}(\p_i A_j-\p_j A_i)^2\right)+\p_t\left(a^3\frac{1}{H}\zeta T^{00}\right)\right)~.
\eeq
In the above expression,  we have dropped a total spatial derivative term. The total time derivative can be removed by a field redefinition, which will however only affect the results at higher order in perturbation theory.
Using $\delta\phi =\sqrt{2\ep}\zeta$ and $\sqrt{2\ep} =- \dot\phi/H$, we have 
\beq\label{id1}
\dot \lambda\zeta = \frac{d \lambda}{d \phi}\frac{d \phi}{d t} \zeta =-\p_\phi\lambda H \delta\phi
\eeq
and with this substitution, the action agrees to leading order in slow-roll with the action in the uniform curvature gauge given in equation (2.23) of \cite{Seery:2008ms} and with equation (30) of \cite{Caldwell:2011ra} in the special case discussed there.

Note that this simple form of the action enables us to generalize our results, in a straightforward manner, to the case where the scalar field $\phi$ is not the inflaton but either an isocurvature field or a curvaton, since the action will take the same simple form in these cases.


\section{The correlation of curvature perturbations with the magnetic fields} \label{crosscorr}

In order to compute the higher-order correlation function during inflation, we adopt a very useful and powerful tool of the in-in formalism \cite{Maldacena:2002vr}. In this formalism, the expectation value of an operator $\mathcal{O}$ at time $\tau_I$ is given by
\begin{equation}
    \label{exp1}
    \lle \Omega \rb \mathcal{O}(\tau_I) \lb \Omega \rgr =\lle
    0\rb
    {\bar \Torder}\left(e^{i\int_{-\infty}^{\tau_I}d\tau \Hint}\right) \mathcal{O}(\tau_I) \Torder \left(e^{-i\int_{-\infty}^{\tau_I} d\tau \Hint}\right)\lb
    0 \rgr
\end{equation}
where $\lb \Omega \rgr$ is the vacuum of the interacting theory, $\lb 0\rgr$ is the vacuum of the free theory,
$\Torder$ and $\bar \Torder$ are time ordering and time anti-ordering operators, respectively, and $\Hint$ is the
interaction Hamiltonian for time $\tau$.  

From the interaction Hamiltonian in (\ref{Hint}) and using the rules of \cite{Giddings:2010ui}, we obtain the cross-correlation of the curvature perturbation with the electromagnetic field
\bea
\left<\zeta(\tau_I,\vect{k}_1)A_i(\tau_I,\vect{k}_2)A_j(\tau_I,\vect{k}_3)\right>&=&-  (2\pi)^3\delta^{(3)}(\vect{k}_1+\vect{k}_2+\vect{k}_3)\nonumber\\
&\times&
\left[\left(\delta_{il}-\frac{k_{2,i}k_{2,l}}{k_2^2}\right)\left(\delta_{lj}-\frac{k_{3,l} k_{3,j}}{k_3^2}\right) \left(\mathcal{I}_1+\vect{k}_2\cdot \vect{k}_3\, \mathcal{I}_2\right)\right.\nonumber\\
& &-\left.\left(\delta_{il}-\frac{k_{2,i} k_{2,l}}{k_2^2}\right){k_{3,l}}\left(\delta_{jm}-\frac{k_{3,j} k_{3,m}}{k_3^2}\right)k_{2,m}\,\mathcal{I}_2\right]
\eea
with the integrals 
\beq
\mathcal{I}_1= 2\,{\rm Im} \left[\zeta_{k_1}(\tau_I)A_{k_2}(\tau_I)A_{k_3}(\tau_I)\int d\tau \tau \lambda'(\tau) \zeta_{k_1}^*(\tau)\p_{\tau}\left(A_{k_2}^*(\tau)\right)\p_{\tau}\left(A_{k_3}^*(\tau)\right)\right]
\eeq
and 
\beq
\mathcal{I}_2=2\,{\rm Im} \left[\zeta_{k_1}(\tau_I)A_{k_2}(\tau_I)A_{k_3}(\tau_I)\int d\tau \tau \lambda'(\tau) \zeta_{k_1}^*(\tau)A_{k_2}^*(\tau)A_{k_3}^*(\tau)\right]
\eeq
where the mode function $\zeta_k(\tau)$ is given in (\ref{zeta}) and the mode function $A_k(\tau)$ is obtained from (\ref{vk}) which is given by
\beq\label{Ak}
A_k(\tau) = \frac{1}{\sqrt{\lambda_I}}\frac{\sqrt{\pi}}{2}e^{i\pi(1+n)/2}\sqrt{-\tau}\left(\frac{\tau}{\tau_I}\right)^nH^{(1)}_{\frac{1}{2}+n}(-k\tau)~.
\eeq
Inserting the expressions for the mode functions, and using the property of the Hankel functions
\beq
\p_x\left(x^n H_n^{(1,2)}(x)\right) = x^n H_{n-1}^{(1,2)}(x)~,
\eeq
we can rewrite the two integrals as 
\bea
\mathcal{I}_1&=&(-2n)~ |\zeta^{(0)}_{k_1}(\tau_I)|^2 |A^{(0)}_{k_2}(\tau_I)| |A^{(0)}_{k_3}(\tau_I)| ~k_2 k_3~\tilde{\mathcal{I}}_n^{(1)}\\
\mathcal{I}_2&=&(-2n)~ |\zeta^{(0)}_{k_1}(\tau_I)|^2 |A^{(0)}_{k_2}(\tau_I)| |A^{(0)}_{k_3}(\tau_I)| ~ \tilde{\mathcal{I}}_n^{(2)}
\eea
where $\zeta^{(0)}_{k}(\tau_I)$ and $A^{(0)}_{k}(\tau_I)$ are the asymptotic super horizon values of the mode functions and are given by
\bea
|\zeta^{(0)}_{k}(\tau_I)| &=& \frac{1}{\sqrt{2\ep}} \frac{H}{\sqrt{2 k^3}}~,\\
|A_k^{(0)}(\tau_I)| &=& \frac{1}{\sqrt{\lambda_I}}\frac{\Gamma(n+1/2)}{\Gamma(3/2)}2^{n-1}\frac{1}{\sqrt{2k}}(-k\tau_I)^{-n}~,
\eea
and the integrals are 
\bea
\tilde{\mathcal{I}}_n^{(1)}&=& \frac{\pi^3}{2}\frac{2^{-2n-1}}{\Gamma^2(n+1/2)}(-k_2\tau_I)^{n+1/2}(-k_3\tau_I)^{n+1/2}\nonumber\\
&\times& {\rm Im} \biggl[ (1+ik_1\tau_I)e^{-ik_1\tau_I}H_{n+1/2}^{(1)}(-k_2\tau_I)H_{n+1/2}^{(1)}(-k_3\tau_I)\Biggr.\nonumber\\
&\times&\biggl. \int^{\tau_I} d\tau \tau(1-ik_1\tau)e^{ik_1\tau}H_{n-1/2}^{(2)}(-k_2\tau)H_{n-1/2}^{(2)}(-k_3\tau)\biggr]~,
\eea
\bea
\tilde{\mathcal{I}}_n^{(2)}&=& \frac{\pi^3}{2}\frac{2^{-2n-1}}{\Gamma^2(n+1/2)}(-k_2\tau_I)^{n+1/2}(-k_3\tau_I)^{n+1/2}\nonumber\\
&\times& {\rm Im} \biggl[ (1+ik_1\tau_I)e^{-ik_1\tau_I}H_{n+1/2}^{(1)}(-k_2\tau_I)H_{n+1/2}^{(1)}(-k_3\tau_I)\biggr.\nonumber\\
&\times&\biggl. \int^{\tau_I} d\tau  \tau(1-ik_1\tau)e^{ik_1\tau}H_{n+1/2}^{(2)}(-k_2\tau)H_{n+1/2}^{(2)}(-k_3\tau)\biggr]~.
\eea
Using $d\ln a = H d t $, one has $-2n = - (\p \lambda_I/\p\ln a)/\lambda_I=-\dot\lambda_I/(H\lambda_I)$, and in this way the correlation function becomes
\bea
\left<\zeta(\tau_I,\vect{k}_1)A_i(\tau_I,\vect{k}_2)A_j(\tau_I,\vect{k}_3)\right>&=&\frac{1}{H}\frac{\dot\lambda_I}{\lambda_I}(2\pi)^3\delta^{(3)}(\vect{k}_1+\vect{k}_2+\vect{k}_3) |\zeta^{(0)}_{k_1}(\tau_I)|^2 |A^{(0)}_{k_2}(\tau_I)| |A^{(0)}_{k_3}(\tau_I)|\nonumber\\
&\times&
\left[\left(\delta_{il}-\frac{k_{2,i}k_{2,l}}{k_2^2}\right)\left(\delta_{lj}-\frac{k_{3,l} k_{3,j}}{k_3^2}\right) \left(k_2 k_3\, \tilde{\mathcal{I}}_n^{(1)}+\vect{k}_2\cdot \vect{k}_3\, \tilde{\mathcal{I}}_n^{(2)}\right)\right.\nonumber\\
& &-\left.\left(\delta_{il}-\frac{k_{2,i} k_{2,l}}{k_2^2}\right){k_{3,l}}\left(\delta_{jm}-\frac{k_{3,j} k_{3,m}}{k_3^2}\right)k_{2,m}\,\tilde{\mathcal{I}}_n^{(2)}\right]~.
\eea
This expression represents our most general result, where the gauge field indices have not yet been contracted.
The cross-correlation of the curvature perturbation with the magnetic fields is now given by
\bea\nonumber
\left<\zeta(\tau_I,\vect{k}_1){\bf B}(\tau_I,\vect{k}_2)\cdot {\bf B}(\tau_I,\vect{k}_3)\right>&=& -\frac{1}{a_0^4}\left(\delta_{ij} \vect{k}_2\cdot \vect{k}_3-\vect{k}_{2,i}\vect{k}_{3,j}\right)\left<\zeta(\tau_I,\vect{k}_1)A_i(\tau_I,\vect{k}_2)A_j(\tau_I,\vect{k}_3)\right>\\ & &
\eea
which leads to the final result
\bea \label{zBB}
\left<\zeta(\tau_I,\vect{k}_1){\bf B}(\tau_I,\vect{k}_2)\cdot {\bf B}(\tau_I,\vect{k}_3)\right> 
&=&-\frac{1}{H}\frac{\dot\lambda_I}{\lambda_I}(2\pi)^3\delta^{(3)}(\vect{k}_1+\vect{k}_2+\vect{k}_3) |\zeta^{(0)}_{k_1}(\tau_I)|^2 |A^{(0)}_{k_2}(\tau_I)| |A^{(0)}_{k_3}(\tau_I)|\nonumber\\
&\times&
\left[\left(\vect k_2\cdot \vect k_3 +\frac{(\vect k_2\cdot \vect k_3)^3}{k_2^2 k_3^2}\right) k_2 k_3 \tilde{\mathcal{I}}_n^{(1)}+2(\vect{k}_2\cdot \vect{k}_3)^2 \tilde{\mathcal{I}}_n^{(2)}\right]~.\nonumber\\
\eea	 
Note that the polarization factor multiplying the integrals is slightly different than in equation (41) of \cite{Caldwell:2011ra}, where it appears that part of the polarization tensor product was missed.  

The solutions of the integrals for different values of $n$ are listed in the appendix. However, for the most interesting case of $n=2$, we find
\bea
\tilde{\mathcal{I}}_2^{(1)}&=& \frac{-1}{(k_2 k_3)^{3/2} k_t^2} \nonumber \\
&\times&\left[-k_1^3 - 2 k_1^2 (k_2 + k_3) - 
 2 k_1 (k_2^2 + k_2 k_3 + k_3^2) - (k_2 + k_3) (k_2^2 + k_2 k_3 + 
    k_3^2)\right]
\eea
and 
\bea \label{I2}
\tilde{\mathcal{I}}_2^{(2)}&=& \frac{-1}{(k_2 k_3)^
  {5/2} k_t^2} \nonumber \\
   &\times&\left[(k_1 + k_2)^2 (-3 k_1^3 - 3 k_1^2 k_2 - 
      k_2^3) + (k_1 + k_2) (-9 k_1^3 - 6 k_1^2 k_2 - 
      2 k_2^3) k_3\right.  \nonumber \\
      &+& (-9 k_1^3 - 6 k_1^2 k_2 - 
      2 k_1 k_2^2 - 2 k_2^3) k_3^2 \nonumber \\
      & -&\left. 2 (2 k_1^2 + k_1 k_2 + k_2^2) k_3^3 - 
   2 (k_1 + k_2) k_3^4 - k_3^5 + 
   3 k_1^3 k_t^2 (\gamma+\ln(-k_t\tau_I) )\right]
\eea
where we have defined $k_t =k_1+k_2+k_3$ and $\gamma$ is the Euler gamma constant. 


\subsection{The flattened shape}

It is interesting to note that the $\ln(-k_t\tau_I)$ term only appears in (\ref{I2}), and can therefore not cancel out in general. This term will be most important when $k_1$ is maximized in the flattened shape with $k_1 =2 k_2=2 k_3$ as already observed in \cite{Caldwell:2011ra}. For length scales relevant for CMB, the logarithm will give an enhancement by a factor 60, but on smaller scales it can be even larger. Since the logarithm completely dominates the integral in the flattened limit, it is easy to estimate the size of $b_{NL}$ in this limit. When the logarithmic term dominates, we have   
 \beq
 \tilde{\mathcal{I}}_2^{(2)} \simeq -\frac{3k_1^3}{(k_2 k_3)^
  {5/2} }  \ln(-k_t\tau_I)~.
\eeq
The contribution of this term to the non-Gaussian cross-correlation function is 
\beq
\left<\zeta(\tau_I,\vect{k}_1){\bf B}(\tau_I,\vect{k}_2)\cdot {\bf B}(\tau_I,\vect{k}_3)\right> \simeq  6\frac{1}{H}\frac{\dot\lambda_I}{\lambda_I}\frac{k_1^3}{(k_2 k_3)^{1/2}}  \ln(-k_t\tau_I) |\zeta^{(0)}_{k_1}(\tau_I)|^2 |A^{(0)}_{k_2}(\tau_I)| |A^{(0)}_{k_3}(\tau_I)|~.
\eeq
For $n=2$, we have $(\dot\lambda_I/H \lambda_I) = 2n=4$, and using $k_1 =2 k_2=2 k_3$ in the flattened limit, we have   
\beq
\left<\zeta(\tau_I,\vect{k}_1){\bf B}(\tau_I,\vect{k}_2)\cdot {\bf B}(\tau_I,\vect{k}_3)\right>\simeq 96  \ln(-k_t\tau_I) P_{\zeta}(k_1)P_{B}(k_2)~.
\eeq
For the reasonable value of the logarithm, corresponding to the largest observable scale today, $ \ln(-k_t\tau_I) \sim -60$, we then obtain in the flattened limit
\beq
\left|b_{NL}^{flat}\right| \sim 5760 ~.
\eeq
This is a quite significant contribution to the non-Gaussianity as compared to the a priori expected level of order unity. 


\subsection{The squeezed limit}

Now let us consider the special limit where the wavelength of the curvature perturbation is much longer than the wavelength of the magnetic fields. In this limit we have $\vect k_1\to 0$ and $\vect k_3 \to -\vect k_2 \equiv -\vect k$. Using the asymptotic behavior for the real and imaginary parts of the Hankel function ${\rm Re}[H^{(1)}_n(x)] \propto x^n$ and ${\rm Im}[H^{(1)}_n(x)] \propto x^{-n}$ for $x\to 0$, it is possible to verify, that in the squeezed limit, the integrals reduce to
\beq\label{besselint}
\tilde{\mathcal{I}}_n^{(1)}=\pi \int^{\tau_I} d\tau \tau J_{n-1/2}(-k\tau)Y_{n-1/2}(-k\tau)
\eeq
and
\beq
\tilde{\mathcal{I}}_n^{(2)}= \tilde{\mathcal{I}}_{n+1}^{(1)}~.
\eeq
In the squeezed limit, the cross-correlation therefore reduces to
\bea
\left<\zeta(\tau_I,\vect{k}_1)A_i(\tau_I,\vect{k}_2)A_j(\tau_I,\vect{k}_3)\right>&=&\frac{1}{H}\frac{\dot\lambda_I}{\lambda_I}(2\pi)^3\delta^{(3)}(\vect{k}_1+\vect{k}_2+\vect{k}_3)|\zeta^{(0)}_{k_1}(\tau_I)|^2 |A^{(0)}_{k}(\tau_I)|^2\nonumber\\
&\times&
\left(\delta_{ij}-\frac{k_i k_j}{k^2}\right)k^2 \left(\tilde{\mathcal{I}}_n^{(1)} - \tilde{\mathcal{I}}_{n+1}^{(1)}\right)
\eea
For integer values of $n$, it can be proven that the integral in (\ref{besselint}) is
\beq\label{simpleint}
\tilde{\mathcal{I}}_n^{(1)}=(n-1/2)/k^2
\eeq
which (in the $\vect k_1 \to 0$ limit) gives 
\bea\label{sqe1}
\left<\zeta(\tau_I,\vect{k}_1)A_i(\tau_I,\vect{k}_2)A_j(\tau_I,\vect{k}_3)\right>&=&-\frac{1}{H}\frac{\dot\lambda_I}{\lambda_I}(2\pi)^3 \delta^{(3)}(\vect{k}_1+\vect{k}_2+\vect{k}_3) \nonumber \\
&\times &\left(\delta_{ij}-\frac{k_i k_j}{k^2}\right)|\zeta^{(0)}_{k_1}(\tau_I)|^2 |A^{(0)}_{k}(\tau_I)|^2\nonumber\\
&=&
-\frac{1}{H}\frac{\dot\lambda_I}{\lambda_I}\left<\zeta(\tau_I, \vect k_1)\zeta(\tau_I,-\vect k_1)\right>\left< A_i(\tau_I,\vect k_2)A_j(\tau_I,\vect k_3)\right>~.
\eea
One can also verify numerically that (\ref{simpleint}) also holds for real non-integer values of $n$.

For the cross-correlation of the curvature perturbations with the magnetic fields, we then obtain for $n>0$
\beq\label{sqe2}
\left<\zeta(\tau_I,\vect{k}_1){\bf B}(\tau_I,\vect{k}_2)\cdot {\bf B}(\tau_I,\vect{k}_3)\right> =-\frac{1}{H}\frac{\dot\lambda_I}{\lambda_I}(2\pi)^3\delta^{(3)}(\vect{k}_1+\vect{k}_2+\vect{k}_3)P_\zeta(k_1) P_B(k_2)~.
\eeq
This agrees with the squeezed limit result in equation (64) of \cite{Caldwell:2011ra}, when using $\dot \lambda\zeta = -\p_\phi\lambda H \delta\phi$ and inserting the specific form of the coupling $\lambda(\phi) = \exp(2\phi/M)$ used there. This agreement is however a coincidence because the difference in the polarization sums noted after  (\ref{zBB}) vanishes in the squeezed limit. In fact, if we had not taken the trace of $B_i B_j$ in the correlation function, the results would no longer agree, even in the squeezed limit.  We also note that both of these results disagree with \cite{Motta:2012rn} in the squeezed limit, which used an interaction Hamiltonian where the leading order term in derivatives of $\lambda$ is a total derivative, which complicates the calculations. This can be seen by comparing our eq.(\ref{Hint}) with eq.(46) of \cite{Motta:2012rn}. In eq.(46) of \cite{Motta:2012rn} the interaction Hamiltonian is proportional to $\lambda$, while in our eq.(\ref{Hint}), we showed that the physical part of the interaction Hamiltonian is proportional only to the derivative of the coupling, $\dot \lambda$, and the term proportional to $\lambda$ is a total derivative, such that the cross-correlation vanishes to leading order in the limit of a constant coupling. 

Upon comparing (\ref{sqe2}) with
\beq
\left<\zeta(\tau_I,\vect{k}_1){\bf B}(\tau_I,\vect{k}_2)\cdot {\bf B}(\tau_I,\vect{k}_3)\right> =b_{NL}^{local}(2\pi)^3\delta^{(3)}(\vect{k}_1+\vect{k}_2+\vect{k}_3)P_\zeta(k_1) P_B(k_2)
\eeq
in the squeezed limit, we obtain the local type magnetic non-linearity parameter
\beq
b_{NL}^{local}= -\frac{1}{H}\frac{\dot\lambda_I}{\lambda_I}~,
\eeq
in agreement with the magnetic consistency relation \cite{Jain:2012ga}.


\section{Discussion and conclusions} \label{diss}

In a model where conformal invariance of a $U(1)$ vector field is broken by a direct coupling of the vector field with the inflaton, we have calculated the three-point  cross-correlation function of two powers of vector field with the primordial curvature perturbation, $\left<\zeta(\vect{k}_1){\bf A}_i(\vect{k}_2){\bf A}_j(\vect{k}_3)\right>$, generated during inflation. We note in passing that the result can be generalized straightforwardly to the case where $\zeta$ is replaced by an isocurvature field or a curvaton, under the identification in (\ref{id1}).

In the case where the vector field is identified with the electromagnetic field, the strength of the cross-correlation is parametrized by the magnetic non-linearity parameter $b_{NL}$ introduced in \cite{Jain:2012ga}. In the squeezed limit where the momentum of the curvature perturbation vanishes, the magnetic non-linearity parameter takes a local form
$b_{NL}^{local}= -(1/H)(\dot\lambda_I/\lambda_I)$~,
in agreement with the magnetic consistency relation \cite{Jain:2012ga}. On the other hand we find some  disagreements with the results of \cite{Motta:2012rn}, even in the squeezed limit. 

With the parametrization of the coupling of the form $\lambda(\phi(\tau))= \lambda_I(\tau/\tau_I)^{-2n}$, the magnetic non-linearity parameter can be related to the spectral index of the magnetic field power spectrum $n_B=4-2n$ for $n>0$, giving the consistency relation $b^{local}_{NL} = n_B-4$. Thus in the most interesting case of a scale invariant magnetic field spectrum $n_B=0$, the non-linearity parameter is non-vanishing  \cite{Jain:2012ga}.

The full shape function is maximal in the flattened shape, where the momentum of the curvature perturbation is maximal and is enhanced by a logarithmic factor, in agreement with\footnote{Note however that we disagree on some details regarding the polarizations sums in eq.(\ref{zBB}).} \cite{Caldwell:2011ra}. For length scales relevant for CMB, the logarithm will give an enhancement by a factor of $60$. Thus, for a flat magnetic field power spectrum, the non-linearity parameter in the flattened limit becomes of order $  |b_{NL}|  \sim  \mathcal O(10^3) $ depending on the scale. 
It will be interesting to understand if such a large $b_{NL}$ could be detectable. Previously it has been argued that the cross-correlation bispectrum of the metric perturbation with the magnetic fields can be observable through the combined survey of large scale structure and Faraday rotation measurements \cite{Stasyszyn:2010kp,Caldwell:2011ra}. We leave it for the future to explore this aspect in more details.

On a different note, the identification of the vector field with the electromagnetic field is problematic from a theoretical point of view due to the strong coupling problem. Although we do not yet have a complete solution to the strong coupling problem, one can consider the model as a simple effective model for primordial inflationary magnetogenesis, assuming that a possible consistent UV complete model might not affect the generic phenomenological features \cite{Kahniashvili:2012vt}.

Finally, we mention that the results obtained here might find applications beyond inflationary magnetogenesis. If the vector field is identified with a vector curvaton non-minimally coupled to the inflaton, then in a scenario where the CMB arises from a mixture of curvaton and inflaton perturbations, the cross-correlation bispectrum of the curvature perturbation with the vector perturbations will be a new non-Gaussian signature in such models. In such a setup,  there is no strong coupling problem, since the vector curvaton can be arbitrarily weakly coupled to the inflaton at the end of inflation.


\acknowledgments
We would like to thank Ruth Durrer, Nemanja Kaloper and Antonio Riotto for useful discussions. RKJ acknowledges financial support from the Swiss National Science Foundation and thanks the CP$^3$-Origins, Centre for Cosmology and Particle Physics Phenomenology at the University of Southern Denmark for hospitality where part of this work was carried out. MSS is supported by a Jr. Group Leader Fellowship from the Lundbeck Foundation.


\section*{Note added}
While writing up this paper, a related paper appeared \cite{Shiraishi:2012xt} with a discussion of the effect on the CMB of the cross-correlation bispectra between the curvature perturbation and the vector modes. The authors of \cite{Shiraishi:2012xt} use the same interaction Hamiltonian as in \cite{Motta:2012rn}, where the leading order term is a total derivative (see our section 3 and the discussion).


\appendix
\section{Integrals}

Here, we evaluate the integrals $\tilde{\mathcal{I}}_n^{(1)}$ and $\tilde{\mathcal{I}}_n^{(2)}$ for different values of $n$. For $n=0$, the integrals are given by
\beq
\tilde{\mathcal{I}}_0^{(1)} =  - \tilde{\mathcal{I}}_0^{(2)} =- \frac{1}{(k_2 k_3)^{1/2} k_t^2} \big(2 k_1+k_2+k_3\big)
\eeq
For $n=1$, the integrals are given by
\bea
\tilde{\mathcal{I}}_1^{(1)}&=& \frac{1}{(k_2 k_3)^{1/2} k_t^2} \big(2 k_1+k_2+k_3\big) \\
\tilde{\mathcal{I}}_1^{(2)}&=& -\frac{1}{(k_2 k_3)^{3/2} k_t^2} \nonumber \\
&\times& \left(-k_1^3 - 2 k_1^2 (k_2 + k_3) - 
 2 k_1 (k_2^2 + k_2 k_3 + k_3^2) - (k_2 + k_3) (k_2^2 + k_2 k_3 + k_3^2)\right)
\eea
For $n=3$, the integrals are given by
\bea
\tilde{\mathcal{I}}_3^{(1)}&=& -\frac{1}{(k_2 k_3)^{5/2} k_t^2} \nonumber \\
&\times& \left[ (k_1+k_2)^2 \left(3 (-1+\gamma) k_1^3-3 k_1^2 k_2-k_2^3\right) \right. \nonumber \\
& + & (k_1 + k_2) \left((-9 + 6 \gamma) k_1^3 - 6 k_1^2 k_2 - 2 k_2^3\right) k_3 \nonumber \\
& + & \left(3 (-3 + \gamma) k_1^3 - 6 k_1^2 k_2 - 2 k_1 k_2^2 - 2 k_2^3\right)  k_3^2 \nonumber \\
& - & 2 (2 k_1^2 + k_1 k_2 + k_2^2) k_3^3 - 2 (k_1 + k_2) k_3^4 -k_3^5 \nonumber \\ 
& + & \left. 3 k_1^3 k_t^2 \ln(-k_t\tau_I)\right]
\eea
\bea
\tilde{\mathcal{I}}_3^{(2)}&=& \frac{1}{2 (k_2 k_3)^{7/2} k_t^2} \nonumber \\
&\times& \left[ (45 k_1^3 k_t^4 + (15 (-1 + \gamma) k_1^7 + 15 (-3 + 2 \gamma) k_1^6 (k_2 + k_3) \right. \nonumber \\
& -&  3 k_1^5 (6 k_2^2 + 5 (5 - 2 \gamma) k_2 k_3 + 6 k_3^2) 
 - k_1^4 (k_2 + k_3) ((-49 + 30 \gamma) (k_2^2 +k_3^2) + 10 k_2 k_3) \nonumber \\
&+&  4 k_1^2 (k_2 + k_3) (3 k_2^4 + 2 k_2^3 k_3 + 4 k_2^2 k_3^2 + 2 k_2 k_3^3 + 3 k_3^4)  \nonumber \\
&+& k_1^3 ((47 - 15 \gamma) (k_2^4+k_3^4)+ (89 - 30 \gamma) (k_2^3 k_3 + k_2 k_3^3) + 6 (16 - 5 \gamma) k_2^2 k_3^2) \nonumber \\
&+& 4 k_1 (k_2^6 + k_2^5 k_3 + k_2^4 k_3^2 + k_2^3 k_3^3 + k_2^2 k_3^4 + k_2 k_3^5 + k_3^6) \nonumber \\
&+& 2 (k_2 + k_3) (k_2^6 + k_2^5 k_3 + k_2^4 k_3^2 + k_2^3 k_3^3 + k_2^2 k_3^4 + k_2 k_3^5 + k_3^6)) \nonumber \\
& + & \left. 15 k_1^3 k_t^2 (k_1^2 - k_2^2 - k_3^2) \ln(-k_t\tau_I)\right]
\eea

\bibliography{rm-refs}
\bibliographystyle{JHEP}
\end{document}